\documentclass[12pt]{article}
\usepackage{graphicx}

\usepackage{times}

\newenvironment{sciabstract}{%
\begin{quote} \bf}
{\end{quote}}

\newcounter{lastnote}
\newenvironment{scilastnote}{%
\setcounter{lastnote}{\value{enumiv}}%
\addtocounter{lastnote}{+1}%
\begin{list}%
{\arabic{lastnote}.}
{\setlength{\leftmargin}{.22in}}
{\setlength{\labelsep}{.5em}}}
{\end{list}}

\title{Universality and Critical Behavior\\ at the Mott transition}

\author
{P.~Limelette$^{1\ast}$, A.~Georges$^{2,1}$, D.~J\'{e}rome$^1$, P.~Wzietek$^{1}$,
P.~Metcalf$^{3}$,J.M.~Honig$^3$\\
\\
\normalsize{$^{1}$Laboratoire de Physique des Solides (CNRS, U.R.A. 8502),}\\
\normalsize{B\^atiment 510, Universit\'e de Paris-Sud, 91405 Orsay, France}\\
\normalsize{$^{2}$Laboratoire de Physique Th\'eorique de l'Ecole Normale Sup\'erieure (CNRS, U.M.R 8549)}\\
\normalsize{24, rue Lhomond, 75231 Paris Cedex 05, France}\\
\normalsize{$^3$ Department of Chemistry, Purdue University,}\\
\normalsize{West Lafayette, IN 47907, USA}\\
\normalsize{$^\ast$To whom correspondence should be addressed; E-mail: limelette@lps.u-psud.fr}
}

\date{}
\begin{document}
\baselineskip24pt
\maketitle


\begin{sciabstract}
We report conductivity measurements of Cr-doped V$_2$O$_3$  using a variable pressure technique.
The critical behavior of the conductivity near the Mott-insulator to metal critical endpoint is investigated
in detail as a function of pressure and temperature.
The critical exponents are determined, as well as the scaling function associated with the equation of state.
The universal properties of a liquid-gas transition are found.
This is potentially a generic description of
the Mott critical endpoint in correlated electron materials.
\end{sciabstract}



\noindent Published as: {\bf Science vol.302 p.89 (October,3rd,
2003)}

\newpage

Since the early recognition by Mott \cite{Mott49,Mott90} that
electron-electron interactions
are responsible for the insulating character of many transition-metal oxides,
extensive research over the last decade has demonstrated the key importance of
this phenomenon for the physics of strongly correlated electron materials.
Outstanding examples \cite{Imada98} are superconducting cuprates, manganites displaying colossal
magnetoresistance, or fullerene compounds.
There are two routes for achieving a metallic state, starting from a Mott insulating
material. The first is to introduce charge carriers by doping. The second,
closely connected to Mott's original ideas, is to reduce the ratio $U/W$ between the
typical strength of local Coulomb repulsion ($U$) and the typical kinetic energy of
the relevant electrons ($W$). This can be achieved in practice, in some materials,
by selected atomic substitutions or by applying pressure. The most widely
studied example \cite{McWhan70,Jaya70,McWhan71,McWhan73,Kuw80}
is Cr-doped Vanadium sesquioxide (V$_{1-x}$ Cr$_{x}$)$_2$O$_3$
which displays a transition from a paramagnetic Mott insulator to a strongly correlated
metal. The transition into the metallic state can be triggered by lowering
temperature (at sufficiently small chromium concentration $x$), by decreasing $x$ or
by increasing pressure $P$
(early studies\cite{McWhan70,Jaya70,McWhan71,McWhan73} have revealed that decreasing concentration
by $\Delta x\sim -0.01$ corresponds to an applied

pressure of $\Delta P\sim 4 \mbox{kbar}$). The transition is first-order, with a significant reduction
of the lattice spacing through the insulator-to-metal transition, indicating a coupling between
electronic and lattice degrees of freedom. The first-order transition line in the $(P,T)$
- or $(x,T)$ -phase diagram ends in a second-order critical endpoint $(P_c,T_c)$.

We report on transport experiments which allow for a precise identification
of the critical behavior associated with this critical endpoint,
a question of fundamental importance in understanding the Mott
transition. Recent theoretical developments have proposed a description
of the critical behavior in simplified purely electronic models (and also
of the crossovers
between distinct transport regimes close to the critical point).
Despite extensive studies of this material, the critical
behavior has not been elucidated so far experimentally.
The key technique used in the present work is to perform conductivity
measurements as a function of continuously varying pressure, at constant
temperature (see ref.~\cite{Jaya70} for an early study).
This  is particularly well adapted to the present situation in which the
transition line in the $(T,P)$ plane is very sharp.
In contrast, the work of Kuwamoto {\it et al.}\cite{Kuw80}
investigated conductivity as a
function of temperature for a discrete set of chromium concentrations $x$.

We use an isopentane liquid pressure cell, and control the
value of pressure with an accuracy of $1~\mbox{bar}$. Conductivity is
measured
at constant regulated temperature with an accuracy  of order $0.1~\mbox{K}$,
as a function of pressure, using a standard four-probe method.
All our measurements were performed on crystals of (V$_{0.989}$ Cr$_{0.011}$)$_2$O$_3$ grown using
the skull-melter technique followed by appropriate annealing \cite{Harrison80}.
The choice of a Cr concentration $x=0.011$ is
such that the sample is on the insulating side of the transition at ambient pressure but that
a moderate pressure (a few kilobars) drives the system into the metallic state (or, alternatively,
a decrease in temperature). This is visible on the data set (Fig. 1A),
which displays the conductivity $\sigma$ as a function of
pressure $P$, for several temperatures in the range $290\mbox{K}<T<485\mbox{K}$.
These data are obtained by decreasing pressure from $P=6\,\mbox{kbar}$ down to ambient
pressure, going from a high-conductivity metallic regime to a low-conductivity
insulating regime. For temperatures smaller than the critical temperature $T_c$ this transition
is discontinuous, with a sudden jump of the conductivity. In order to locate precisely
this critical point and to demonstrate the first-order nature of the transition, we have performed
hysteresis experiments in which the conductivity is measured during increasing and
decreasing pressure sweeps at a slow rate of order 25 bar/min (Fig. 1A).
From the difference between the measured conductivities
in these two sweeps (Fig. S2), two characteristic pressures can be identified, $P_M(T)$ and $P_I(T)$
($P_M<P_I$), corresponding respectively to the lowest pressure at which
a metallic state can be sustained while decreasing pressure ($P_M$), and to the highest pressure at
which an insulating state can be sustained while increasing pressure ($P_I$). These two spinodal lines
, plotted as a function of temperature on Fig. 1B, merge at the critical
endpoint $(P_c,T_c)$. We can then estimate: $P_c\simeq 3738~\mbox{bar}$, $T_c\simeq 457.5~\mbox{K}$.
Varying pressure rather than temperature is essential for a precise determination of
$T_c$, which is compatible with the early estimate ($\simeq 450\mbox{K}$) by Kuwamoto et al.\cite{Kuw80}.
At the critical temperature, the pressure dependence of $\sigma(P,T_c)$ becomes singular,
with a vertical tangent at the critical pressure $P=P_c$ (Fig. 2A).
For $T>T_c$, this singular behavior is replaced by a continuous variation of the
conductivity with pressure, which nevertheless defines a sharp crossover line in the
$(P,T)$ phase diagram (as also depicted in Fig. 1B). This crossover line
extrapolates to a temperature of order $\sim 500\mbox{K}$ for the pressure
($\sim 5 \mbox{kbar}$) corresponding to the pure V$_2$O$_3$ compound. Interestingly,
the location of this crossover coincides with the one detected in early NMR
experiments\cite{Kerlin73}.

We now show that the critical singularities found in the vicinity of the
critical endpoint $(P_c,T_c)$ can be analyzed in the framework of the scaling theory
of the liquid-gas transition of classical systems\cite{Kad67}.
The analogy between the latter and the finite-temperature
Mott transition has been emphasized early on by Castellani {\it et al.}\cite{Cast79}
(see also Ref.~\cite{Jaya70}).
The insulating phase (in which the Vanadium is mainly in the $\mbox{V}^{3+}$
state, corresponding to the $d^2$ configuration) can be pictured as a ``gas'' phase
with a low density of double occupancies or holes (corresponding to $\mbox{V}^{2+}$ and
$\mbox{V}^{4+}$, or $d^3$ and $d^1$, respectively). The metallic phase corresponds to a
``liquid'' with a sizeable density of holes and double occupancies.
Recently, this analogy has been
given firm theoretical foundations within the framework of a
Landau theory\cite{Kot00,Roz99} derived from dynamical mean-field theory (DMFT)\cite{Georges96}.
In this framework, a scalar order parameter $\phi$ is associated with
the low-energy electronic degrees of freedom which build up the quasiparticle
resonance in the strongly correlated metallic phase close to the transition.
This order parameter couples to the singular part of the double occupancy (hence
providing a connection to the picture described above), as well as to other
observables such as the Drude weight or dc-conductivity.
Because of the scalar nature of the order parameter, the transition falls in
the Ising universality class. Coupling to lattice degrees of freedom can also be
included in the theory\cite{Maj94} without changing this conclusion.
In the following  we denote by $r$ the scaling variable corresponding to the temperature
scaling axis in the Ising model analysis (i.e. to the term $r\phi^2$ in the Landau functional)
and by $h$ the scaling axis corresponding to magnetic field (i.e. to the symmetry-breaking
term $-h\phi$). These scaling variables are a priori linear combinations of $(T-T_c)/T_c$
and $(P-P_c)/P_c$. However, our data are compatible with no or little
mixing, so that we choose in all the following:
$r=(T-T_c)/T_c+\cdots\,,\,h=(P-P_c)/P_c+\cdots$ (the dots indicate higher order terms).
Denoting by $\sigma_c=\sigma(P_c,T_c)$ ($\simeq 449.5\,~\Omega^{-1}\mbox{cm}^{-1}$)
the measured conductivity
at the critical point, it is expected that $\sigma(P,T)-\sigma_c$ depends linearly on the
order parameter $\langle\phi\rangle$ close to the critical point. (This can be
explicitly proven in the context of DMFT).
At $T=T_c$, this implies a critical singularity of the form:
$\sigma(P,T_c)-\sigma_c \sim h^{1/\delta}$ with $\delta$ the critical exponent associated
with the singular dependence of the magnetization at the critical point in the Ising
model. The data in Fig. 2A are very well fitted by this form, as demonstrated in the inset.
Over more than two decades in $h\propto(P-P_c)/P_c$, we find the best-fit value of
the exponent to be $\delta\simeq 3$, i.e. the mean-field value. In a narrow pressure
interval ($\Delta P\simeq 10~\mbox{bar}$) close to the critical pressure,
indication for a crossover towards a value $\delta\simeq 5$ is found,
close to the three-dimensional (3D) Ising value $\delta \simeq 4.814$.

We now address the critical behavior away from $T_c$ by studying the temperature dependence of
the conductivity in the following manner. For $T<T_c$, we focus on
 the conductivity of the metallic state, at the high-pressure
boundary of the coexistence region. That is, we consider
$\sigma^*(T)\equiv\sigma_{met}(P_I(T),T)-\sigma_c $
with $P_I(T)$ the spinodal of the insulating phase.
This quantity, plotted in Fig. 2B, is expected
to display the critical behavior of the order parameter, by analogy with the
liquid-gas transition: $\sigma^*(T) \sim (-r)^\beta$ with $r\propto (T-T_c)/T_c$.
As shown in the inset (Fig. 2B), a mean-field value of the critical exponent
$\beta\simeq 0.5$ is found to fit the data over almost two decades away from the critical
point. In a narrow temperature interval $\Delta T\simeq 4\mbox{K}$ close to $T_c$
($\Delta T/T_c\simeq 0.01$), some indication for a crossover towards a non mean-field
value $\beta\simeq 0.34$ is found, close to the 3D Ising value $\beta\simeq0.327$.
We also consider the derivative of the conductivity with respect to pressure, in the
metallic state, taken on the
same spinodal line: $\chi(T)\equiv (d\sigma_{met}(P,T)/dP)|_{P=P_I(T)}$. This quantity can
be defined as well for $T>T_c$ by taking the derivative at the inflection point of the
$\sigma(P)$ curve (Fig. 1A). Following the liquid-gas analogy, it corresponds to
the magnetic susceptibility in the equivalent Ising model:
$\chi\propto d\langle\phi\rangle/dh$. As shown in Fig. 2C, it is found
to diverge as $\chi \sim \chi_{+}/(T-T_c)^\gamma$ for $T>T_c$ and as
$\chi \sim \chi_{-}/(T_c-T)^{\gamma'}$ for $T<T_c$. The exponent $\gamma$,
as well as the (universal) amplitude ratio $\chi_{+}/\chi_{-}$, are found to
be close to their mean-field values: $\gamma=\gamma'=1$, $\chi_{+}/\chi_{-}= 2$.
Very close to $T_c$, the noise in the numerical derivative involved in
the determination of $\chi$ prevents a reliable determination of deviations from mean-field,
in contrast to the above study of the conductivity itself.

Finally, we demonstrate that the whole set of conductivity data in the metallic phase
can be scaled onto a universal form, which can be written as:
\begin{equation}
\nonumber
\sigma_{met}(P,T)-\sigma_c = (\delta h)^{1/\delta}\,
f_{\pm}\left( \frac{\delta h}{|r|^{\gamma\delta/(\delta-1)}} \right)
\end{equation}
In this expression, $\delta h=h-h_I$ denotes the difference between the ''field''
$h=(P-P_c)/P_c$ and its value on the spinodal line of the insulating phase
$h_I=(P_I-P_c)/P_c$, i.e.: $\delta h=(P-P_I(T))/P_c$.
This amounts to a simple shift of the field variable on the standard form \cite{Kad67}
of the universal equation of states near a liquid-gas critical point.
The functions $f_{+}$ and $f_{-}$ are universal scaling functions
which apply for $T>T_c$ ($r>0$) and $T<T_c$ ($r<0$), respectively.
When written in this form, the equation of state is such that the
order parameter $\sigma^*(T)$ defined above is recovered when the
limit $\delta h\rightarrow 0$ is taken in the right-hand side of Eq.~1.
The pressure-dependent data sets for many different temperatures have
been plotted in this manner (Fig. 3), in which the two exponents $\gamma$ and
$\delta$ were taken as adjustable parameters (Fig. S3) in order to obtain the best
collapse of all the data points onto single curves. This leads to values of
these exponents close to the mean-field ones $\gamma\simeq 1, \delta\simeq  3$,
which provides a strong check on the individual determination of
each critical exponent performed above. The quality of the scaling is seen to
be excellent over a very large range of variation of the scaling variables (several
decades). It is apparent that the scaling functions obey the expected
asymptotic behaviors:
$f_{+}(x\ll 1)\sim x^{1-1/\delta}$, $f_{-}(x\ll 1)\sim x^{-1/\delta}$ and
$f_{\pm}(x\gg 1)\sim\mbox{const.}$. This finding is essential to ensure that
Eq.~1 be compatible with the critical behavior of the order parameter $\sigma^*$ at small
and large field, for both $T<T_c$ and $T>T_c$, investigated previously
in Fig. 2. It also implies that the critical exponents obey the relation
$\gamma=\beta(\delta-1)$, in agreement with the above determination of $\beta$.

These universal scaling properties of the pressure- and temperature-dependent
conductivity experimentally demonstrate that
the electronic degrees of freedom undergo a liquid-gas phase transition
at the Mott critical endpoint. Critical exponents and a universal scaling
function have been determined. Our results are consistent with mean-field values
over most of the parameter range, with some indication for three-dimensional
Ising behavior very close to the transition.
A possible explanation for why the
range of validity of mean-field theory is so large can be put forward by
analogy with the theory of conventional superconductors. There, the key point
is the existence of a very large length scale (the pair coherence length), much
larger than the lattice spacing (or the Fermi wavelength). Here, the Mott insulator
can be thought of as a state in which holes and doubly occupied sites form bound states
due to their Coulomb interaction. The spatial extension $\xi$ of these bound states is related
to their energy (the Mott gap $\Delta$) by $\Delta \sim h^2/(2m\xi^2)$. Given the
measured value of $\Delta$ in samples close to the transition, this leads to the
conclusion that $\xi$ is indeed a large length-scale, of order a few nanometers.
Finally, we emphasize that our results provide experimental
support to the early idea of Ref.~\cite{Cast79}
and to recent theories of the Mott critical endpoint based on the dynamical
mean-field (DMFT) approach\cite{Kot00,Roz99,Georges96}.
While further effort should be devoted to the inclusion of lattice
degrees of freedom in these theories, simplified treatments of these
effects\cite{Maj94} do emphasize the key role of
electronic degrees of freedom in the transition.

\bibliographystyle{Science}


\begin{scilastnote}
\item We acknowledge fruitful discussions with R.~Chitra, S.~Florens,
G.~Kotliar, H.R.~Krishnamurthy, M.~Rozenberg and A.~J.~Millis.
\end{scilastnote}

\clearpage
\begin{center}
Figure captions
\end{center}

\noindent  Fig. 1A: Conductivity as a function of decreasing pressure,
for temperatures ranging from $T=485 \mbox{K}$ ($>T_c= 457.5 \mbox{K}$, orange curves) down
to $T=290 \mbox{K}$ ($<T_c$, blue curves). The dark yellow curve is the conductivity at $T_c$.
Only a selected set of values of $T$ has been displayed, for clarity.
(For a two-dimensional plot of the data, see Fig. S1)
Examples of an hysteresis cycle are shown for $T=290 \mbox{K}$ and $T=348 \mbox{K}$.
For a plot of the difference of conductivities measured in decreasing and increasing pressure
sweeps, see Fig. S2.\\
Fig. 1B: Phase diagram of Cr-doped Vanadium Sesquioxide V$_{1-x}$ Cr$_{x}$)$_2$O$_3$
as a function of pressure and temperature, in the range $1\mbox{bar}<P<6\mbox{kbar}$ and
$290\mbox{K}<T<500\mbox{K}$ investigated in this work. At a given temperature $T$, the metallic state
can be obtained for pressures higher than the spinodal pressure $P_M(T)$, and the insulating state
for pressures lower than the spinodal pressure $P_I(T)$. These two spinodal lines delimit a
pressure range $P_M<P<P_I$ in which the two states coexist (hatched region on the figure).
This coexistence region closes at the critical endpoint $(P_c,T_c)$ ($P_c\simeq 3738 \mbox{bar},
T_c \simeq 457.5 \mbox{K}$). The crossover line above this point (dashed) corresponds to the
inflection point in the $\sigma(P)$ curves.\\
Fig. 1C: Schematic global phase diagram of Cr-doped Vanadium Sesquioxide V$_{1-x}$ Cr$_{x}$)$_2$O$_3$
as a function of pressure and temperature, deduced from Ref.~\cite{McWhan73}.\\

\noindent
Fig.2 These plots demonstrate how the critical exponents $\delta$,
$\beta$ and $\gamma$ can be inferred from the study of the conductivity and of its
derivative with respect to pressure (see text).\\
Fig. 2A: At the critical temperature $T=T_c$, the conductivity $\sigma $
is plotted as a function of pressure. The (plain) red line is a fit to
$\sigma-\sigma_c\sim (P-P_c)^{1/\delta}$, with $\delta=3$. The use of a logarithmic scale
(inset) confirms this value, and also reveals a non mean-field regime for $P$ close to $P_c$.\\
Fig. 2B: Order parameter $\sigma^*(T)=\sigma(P_I(T),T)-\sigma_c$ {\it vs.}
$T/T_c$, for $T<T_c$. The line is a fit to $(T_c-T)^\beta$ with $\beta=0.5$. The inset
(logarithmic scale) reveals a non mean-field regime close to $T_c$.\\
Fig. 2C: Derivative of the conductivity (analogous to a susceptibility $\chi$,
as described in text), for $T<T_c$ and $T>T_c$. The plain lines are fits
to $\chi_{\pm}|T-T_c|^{-\gamma}$, with $\gamma=1$ and $\chi_+/\chi_{-}=2$.

\noindent Fig. 3: Scaling plot of the conductivity onto a universal equation of state.
The whole data set in the metallic state has been used in order to plot
$(\sigma-\sigma_c)/(P-P_I)^{1/\delta}$
{\it vs.} $(P-P_I)/(T-T_c)^{\gamma\delta/(\delta-1)}$, as described in the text.
The data collapse onto two universal curves for $T>T_c$ and $T<T_c$,
corresponding to the universal scaling functions $f_{\pm}$ in Eq.(1).

\clearpage
\begin{figure}[htbp]
\begin{center}
\centerline{\includegraphics[width=\textwidth]{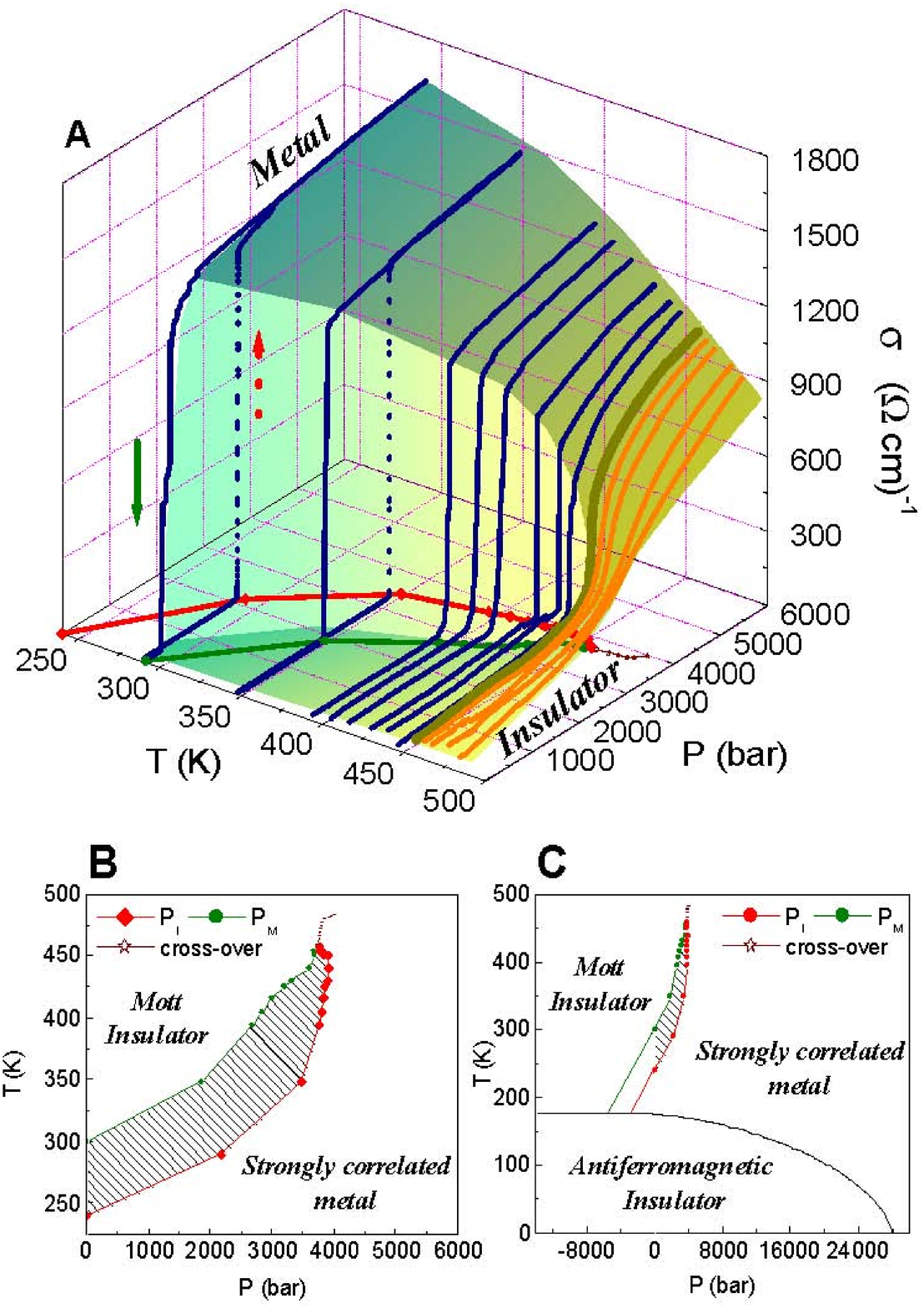}}
\caption{}
\end{center}
\end{figure}

\clearpage

\begin{figure}[htbp]
\begin{center}
\centerline{\includegraphics[width=\textwidth]{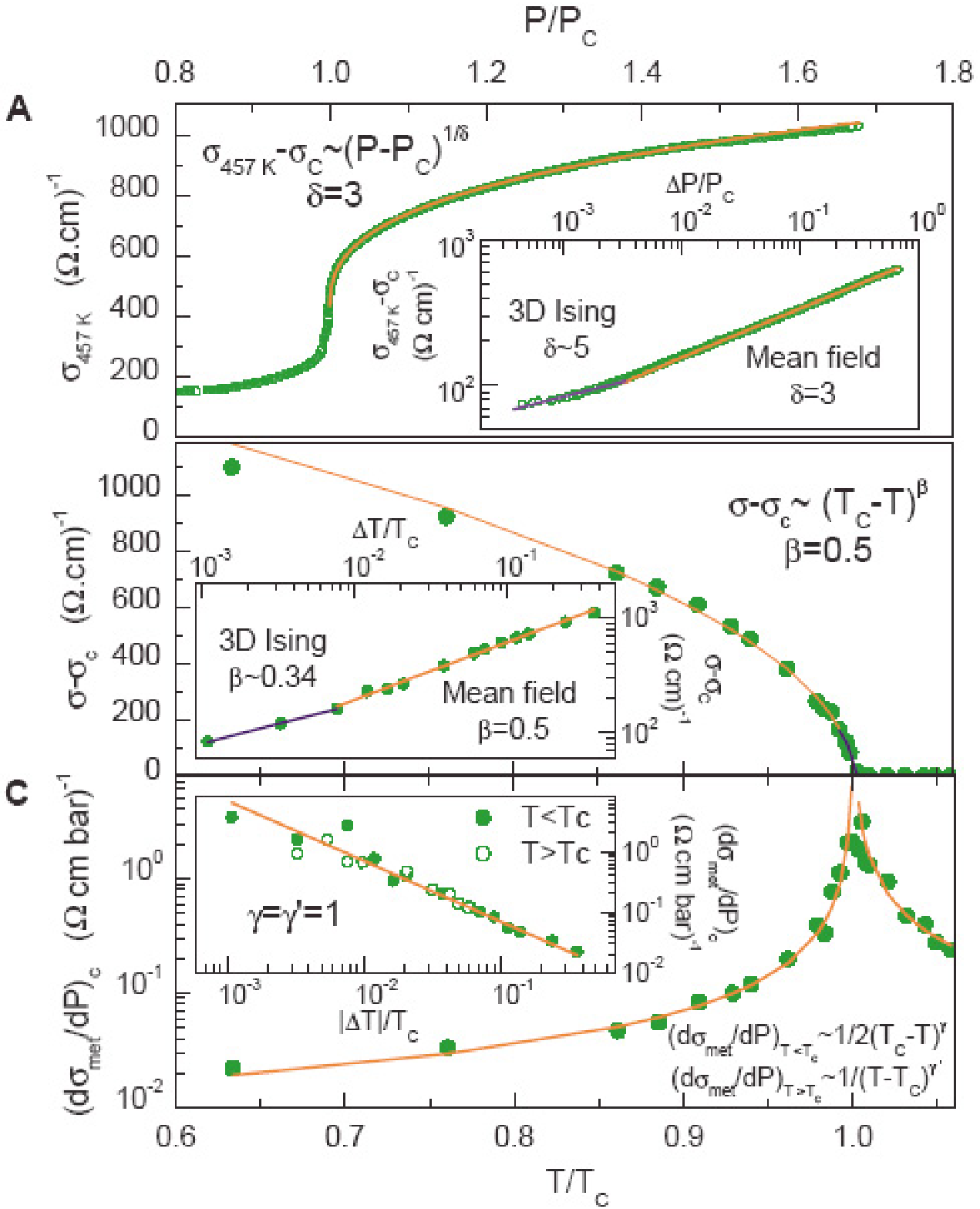}}
\caption{}
\end{center}
\end{figure}

\clearpage

\begin{figure}[htbp]
\begin{center}
\centerline{\includegraphics[width=\textwidth]{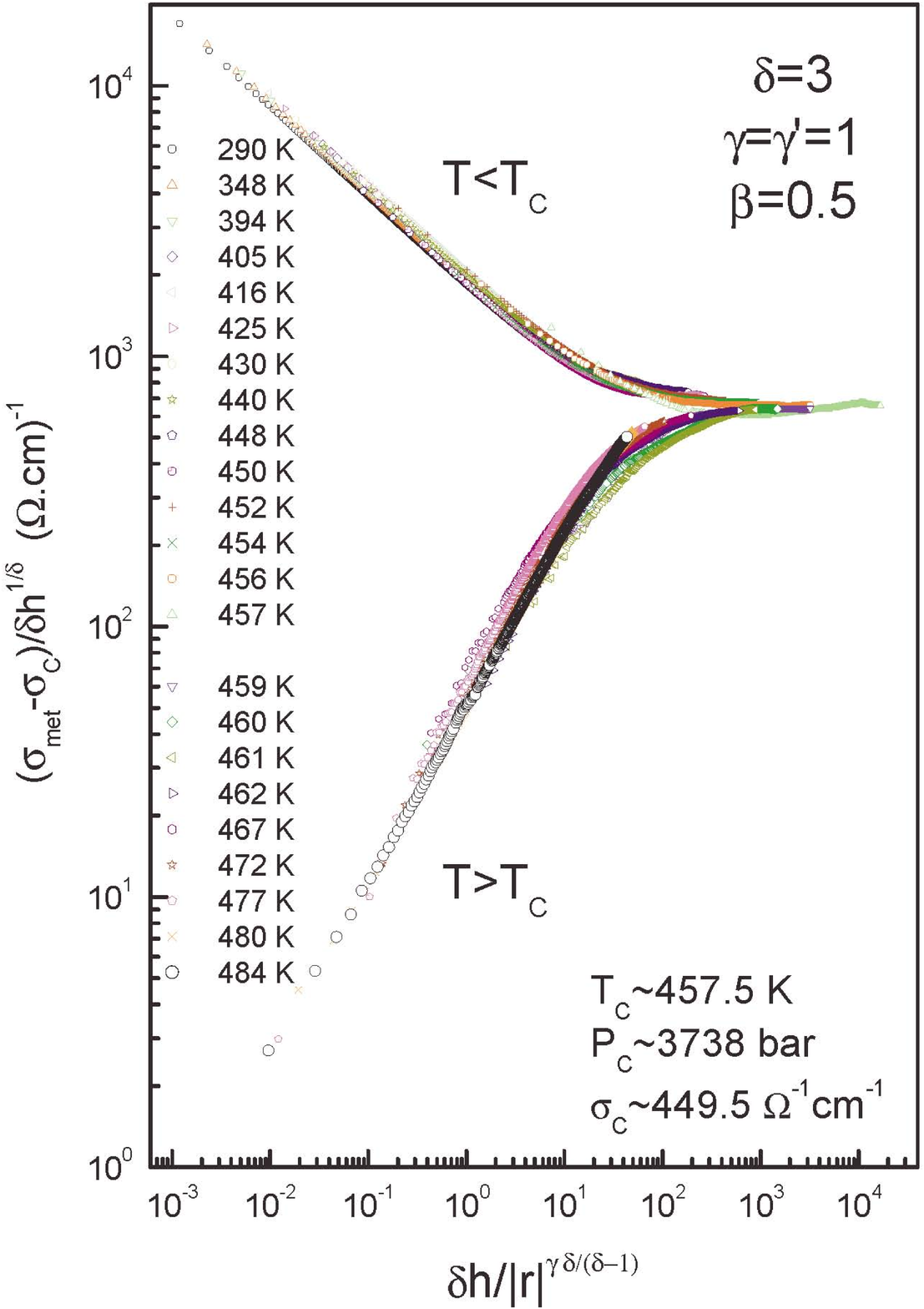}}
\caption{}
\end{center}
\end{figure}

%

\end{document}